\documentclass[aps,prl,twocolumn,showpacs,superscriptaddress,floatfix]{revtex4}
\usepackage[dvips]{graphicx}
\usepackage[english]{babel}
\usepackage{color,enumerate,amsmath,amssymb,amsbsy,amscd,bm}
\usepackage{epsf,epsfig}
\usepackage{epstopdf}

\newcommand{\ket}[1]{|{#1} \rangle}  

\begin{document}
\preprint{}
\title[]{Resonant Adiabatic Passage with Three Qubits}
\author{Sangchul Oh}
\affiliation{Department of Physics, University at Buffalo, The State University of New York,
Buffalo, New York 14260-1500, USA}
\author{Yun-Pil Shim}
\affiliation{Department of Physics, University of Wisconsin-Madison, Madison, Wisconsin 53706, USA}
\author{Jianjia Fei}
\affiliation{Department of Physics, University of Wisconsin-Madison, Madison, Wisconsin 53706, USA}
\author{Mark Friesen}
\affiliation{Department of Physics, University of Wisconsin-Madison, Madison, Wisconsin 53706, USA}
\author{Xuedong Hu}
\affiliation{Department of Physics, University at Buffalo, The State University of New York,
Buffalo, New York 14260-1500, USA}
\date{\today}
\begin{abstract}
We investigate the non-adiabatic implementation of an adiabatic quantum teleportation 
protocol, finding that perfect fidelity can be achieved through resonance. We clarify 
the physical mechanisms of teleportation, for three qubits, by mapping their dynamics 
onto two parallel and mutually-coherent adiabatic passage channels. By transforming into 
the adiabatic frame, we explain the resonance by analogy with the magnetic resonance of 
a spin-1/2 particle. Our results establish a fast and robust method for transferring 
quantum states, and suggest an alternative route toward high precision quantum gates.
\end{abstract}
\maketitle

Fault-tolerant quantum computation requires high-precision quantum gates with noise 
thresholds between $10^{-4}$ and $10^{-2}$, depending on the fault-tolerance 
scheme~\cite{Shor,Preskill}. This stringent requirement poses significant technical 
challenges, even for the more mature qubit architectures, such as those based on trapped 
ions~\cite{Lanyon_Science11}. Identifying gate protocols that are both fast and robust 
is therefore an important research objective for quantum information processing.

One potential approach to robust quantum gates is based on the adiabatic principle -- 
a fundamental tenet of quantum mechanics~\cite{Messiah}. According to the adiabatic 
theorem, a quantum system in an eigenstate remains there, provided that the Hamiltonian 
varies slowly in time. Applications of the adiabatic theorem include the Born-Oppenheimer 
approximation and the Landau-Zener-St\"uckelberg transition at an avoided crossing, 
the latter having been demonstrated in both superconducting and spin 
qubits~\cite{Oliver,Gaudreau12}.  Other experimental implementations include adiabatic 
population transfers between two or three-level systems, known as adiabatic passages 
(AP)~\cite{Oreg84,Bergmann98,Kral07,Shore08}, which have been demonstrated in atomic, 
molecular, and optical devices. There are also theoretical proposals for realizing AP 
with superconducting qubits~\cite{Norigroup} and quantum dot arrays~\cite{Greentree04}. 
Adiabatic quantum information processing~\cite{Farhi01} entails the adiabatic transformation 
of the ground state of an initial Hamiltonian into that of a target Hamiltonian. 
Compared to the quantum circuit model, adiabatic gates are resistant to decoherence when 
a finite excitation gap persists throughout the evolution, and they are robust to gating 
errors, by virtue of adiabaticity. This can be a drawback however, since the maximum
speed of an adiabatic gate is also proportional to the spectral gap.

In this Letter we investigate a non-adiabatic form of adiabatic quantum teleportation (AQT).  
Conventional AQT was proposed in the context of fault tolerant quantum 
computation~\cite{Bacon09}. Here, we focus on systems with three qubits, where we can solve 
the evolution analytically.  We show that resonances occur, enabling teleportation that 
is fast, fault-tolerant, and potentially perfect. This surprising effect can be explained 
in the language of spin resonance, by transforming into the adiabatic frame. Our results 
point toward a new paradigm for quantum algorithms, based on fast adiabatic gates. Our work 
also provides an interesting mapping between three coupled qubits and a three-level atom, 
which could lead to further spin analogies from atomic $\Lambda$-system physics. 
The experimental requirements for implementing AQT have already been demonstrated in the 
laboratory for triple quantum dots~\cite{Gaudreau12} and superconducting 
circuits~\cite{Martinis11}. Our results could therefore be tested immediately. 
\begin{figure}[htbp]
\includegraphics[scale=0.9,angle=0]{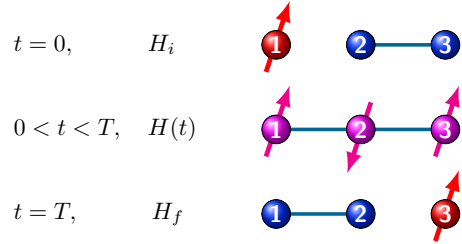}
\caption{(color online). The couplings and the information distribution (by color and 
arrow) among three qubits are shown at the initial, intermediate, and final stages of 
adiabatic quantum teleportation.}
\label{Fig1}
\end{figure}

The adiabatic quantum teleportation protocol is illustrated in Fig.~\ref{Fig1}. Initially, 
qubit~1 is isolated and prepared in an arbitrary superposed state, while qubits 2 and 3 are 
coupled, as described below, and prepared in the maximally entangled singlet state. The
antiferromagnetic coupling between qubits 1 and 2 (2 and 3) is then turned on (off) slowly. 
When the evolution is complete, the quantum state of qubit 1 will be teleported to qubit 3. 
As proposed in Ref.~\cite{Bacon09}, the scheme succeeds when the run time $T$ satisfies 
the adiabatic theorem.

To explore non-adiabatic effects, we solve the exact dynamics of the three-qubit system. 
It is governed by a time-dependent Hamiltonian, which smoothly changes from the initial 
Hamiltonian $H_i$ at $t=0$ to the final Hamiltonian $H_f$ at $t=T$:
\begin{align}
H(t) = f(t)\, H_i + g(t)\, H_f\,.
\label{Eq:Hamiltonian}
\end{align}
The initial and final Hamiltonians are given by
\begin{subequations}
\begin{align}
H_i &= J\left(\,    \sigma_{2x}\,\sigma_{3x} + \sigma_{2y}\,\sigma_{3y}
            + \gamma\,\sigma_{2z}\,\sigma_{3z}  \,\right)\,,\\
H_f &= J\left(\,    \sigma_{1x}\,\sigma_{2x} + \sigma_{1y}\,\sigma_{2y}
            + \gamma\,\sigma_{1z}\,\sigma_{2z}  \,\right)\,,
\end{align}
\end{subequations}
where $\sigma_{i\mu}$ are the Pauli operators with $i=1,2,3$ and $\mu=x,y,z$,  and $J$ is 
the strength of the qubit-qubit coupling. The anisotropy parameter $\gamma = 0$ corresponds 
to an {\it XX} coupling, available in superconducting qubits, while $\gamma =1$ corresponds 
to the isotropic Heisenberg coupling of spin qubits. The interpolation or switching functions 
$f(t)$ and $g(t)$ satisfy $f(0) = g(T) =1$ and $f(T) = g(0) = 0$. Here we consider two ways 
to connect $H_i$ to $H_f$: (i) a linear interpolation with $f(t) = 1 -t/T$ and $g(t)= t/T$, 
and (ii) a harmonic interpolation with $f(t)=\cos({\pi t}/{2T})$ and 
$g(t) = \sin({\pi t}/{2T})$.

AQT begins with an initial three-qubit state given by
\begin{align}
\ket{\psi(0)} = \bigl(a\ket{0} + b\ket{1}\bigr)_1
\otimes \frac{1}{\sqrt{2}}\bigl(\ket{01} -\ket{10}\bigr)_{2,3}\,,
\label{Eq:initial_state}
\end{align}
where $a\ket{0}+b\ket{1}$ is the arbitrary state to be teleported. The success of AQT is 
measured by the fidelity $F(T) = |\langle{\psi_T}|{\psi(T)}\rangle|^2$ where $\ket{\psi(T)}$ 
is the final state at time $T$ and $\ket{\psi_T} = \frac{1}{\sqrt{2}}
\bigl(\ket{01} -\ket{10}\bigr)_{1,2} \otimes \bigl(a\ket{0} + b\ket{1}\bigr)_3$ is 
the target state. The dynamics of AQT is governed by the time-dependent
Hamiltonian~(\ref{Eq:Hamiltonian}), with the initial state~(\ref{Eq:initial_state}).

Hamiltonian~(\ref{Eq:Hamiltonian}) satisfies the commutation relation $[H(t), S_z] = 0$, 
so that the $z$-component of the total spin angular momentum 
$S_z \equiv \frac{1}{2}\left(\sigma_{1z} + \sigma_{2z} + \sigma_{3z}\right)$ is 
a good quantum number, which is conserved during evolution. The three-qubit 
Hamiltonian~(\ref{Eq:Hamiltonian}) is thus block-diagonal:
\begin{align}
H(t) = \underbrace{H_3}_{\rm up}\oplus\underbrace{H_3}_{\rm down}\oplus H_1\oplus H_1\,.
\end{align}
The two $H_1$ operators act on $\ket{000}$ and $\ket{111}$, while the two $H_3$ operators 
act on the distinct subspaces ${\cal H}_{1/2} = \text{Span}(\ket{100}, \ket{010}, \ket{001})$ 
and ${\cal H}_{-1/2} =\text{Span}(\ket{011},\ket{101},\ket{110})$, and have the same form
\begin{align}
H_3(t) = J\begin{bmatrix}
(f-g)\gamma  &2g           & 0     \\
2g           &-(f+g)\gamma & 2f    \\
0            &2f           & -(f-g)\gamma
\end{bmatrix}\,.
\label{Eq:Hamiltonian_3}
\end{align}
Interestingly, $H_3$ is also the AP Hamiltonian for a 3-level 
atom~\cite{Bergmann98,Kral07,Shore08}, with the switching functions $f(t)$ and $g(t)$ being 
the Stokes and pump pulses in the context of AP. For the initial states we consider, 
the $H_1$ operators are never involved in the system evolution. The AQT protocol therefore 
consists of two parallel, identical and mutually-coherent APs governed by $H_3$, corresponding 
to the $S_z = \pm \frac{1}{2}$ components of the three-qubit system.

To understand the dynamics of AQT, we solve the time-dependent Schr\"odinger equation with 
Hamiltonian (\ref{Eq:Hamiltonian_3}) in two ways. First, we consider the adiabatic limit, 
for which there is a mapping between AQT and two  mutually-coherent APs. Second, we obtain 
numerical solutions (and in one case, an analytical solution) for finite $T$. We also 
consider the separate cases of {\it XX} and Heisenberg couplings.

The adiabatic theorem states that, starting from an eigenstate $\ket{E_n(0)}$, the 
adiabatically evolved state 
$\ket{\psi(t)} \simeq e^{-i/\hbar\int_0^t E_n(t')\,dt' +\gamma_B} \,\ket{E_n(t)}\,$ 
is simply an instantaneous eigenstate, up to a phase factor. Here, $\gamma_B$ is
the Berry phase, and $E_n$ and $\ket{E_n}$ are the instantaneous eigenvalues and 
eigenstates of $H_3(t)$, defined by
\begin{align}
H_3(t)\,\ket{E_n(t)} = E_n(t)\,\ket{E_n(t)}\,.
\label{Eq:instantaneous}
\end{align}

Solving Eq.~(\ref{Eq:instantaneous}) for an {\it XX} coupling ($\gamma=0$), gives 
the instantaneous energy levels $E_0(t) = 0$ and $E_\pm(t) = \pm 2J\sqrt{f^2 + g^2}$, 
with the corresponding eigenstates
\begin{align}
\ket{E_0(t)} &=
   \begin{bmatrix}
   \cos\theta\\
   0\\
   -\sin\theta
   \end{bmatrix}\,,\quad
\ket{E_\pm(t)} =\frac{1}{\sqrt{2}}
\begin{bmatrix}
\sin\theta\\
\pm 1\\
\cos\theta
\end{bmatrix} \,.
\label{Eq:XX_eigenstates}
\end{align}
Here, the mixing angle $\theta=\tan^{-1}[g(t)/f(t)]$ runs from $0$ to $\pi/2$ as time $t$ 
goes from 0 to $T$. For the Heisenberg coupling ($\gamma =1$), the instantaneous energy 
levels are $E_0(t)/J = (f+g)$ and $E_\pm(t)/J = [-f-g\pm 2\sqrt{f^2 -fg + g^2}]$, with 
the corresponding eigenstates
\begin{subequations}
\label{Eq:Heisen_eigenstates}
\begin{align}
\hspace{-4pt} \ket{E_0(t)} &= \frac{1}{\sqrt{3}}
\begin{bmatrix}
1 & 1 & 1
\end{bmatrix}^T\,,\\
\hspace{-4pt} \ket{E_\pm(t)} &=\frac{1}{\sqrt{\cal N}}
\begin{bmatrix}
  \sin\theta\\
  -\cos\theta \pm \sqrt{1-\cos\theta\sin\theta}\\
  \cos\theta -\sin\theta \mp \sqrt{1 -\cos\theta\sin\theta}
 \end{bmatrix},
\end{align}
\end{subequations}
where ${\cal N} \equiv 2(2\cos\theta -\sin\theta)\sqrt{q}+ 4q$, $q = 1-\cos\theta\sin\theta$.

\begin{figure}[htbp]
\includegraphics[scale=0.9,angle=0]{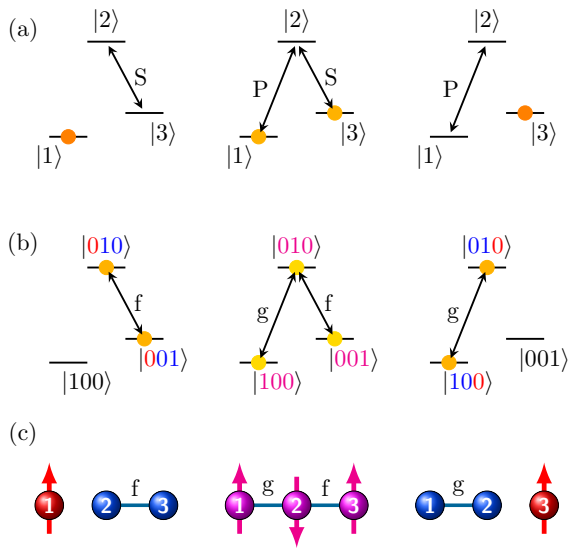}
\caption{(color online). Adiabatic passage protocols for (a) $\Lambda$-type levels, (b)
the ``up" component (red $\ket{0}$) of a 3-qubit system, and (c) the corresponding spin 
configurations.  From right to left: the initial, intermediate, and final stages of 
evolution. In (a) and (b), filled circles represent populations of levels. $S$ and $P$ 
stand for the Stokes and pump pulses. }
\label{Fig:adiabatic_passage}
\end{figure}

Equation~(\ref{Eq:Hamiltonian_3}) governs the evolution of both AQT and conventional AP, 
as depicted in Fig.~\ref{Fig:adiabatic_passage}. For AP, the population of a $\Lambda$-type 
system is transferred from state $\ket{1}$ to $\ket{3}$, while state $\ket{2}$ remains
unpopulated~\cite{Bergmann98,Kral07,Shore08}. Paradoxically, the AP pulse sequence appears 
to occur in reverse order (S followed by P), as shown in panel (a). The instantaneous 
eigenstate used in this evolution is $\ket{E_0(t)}$ from Eq.~(\ref{Eq:XX_eigenstates}). 
For AQT, on the other hand, the instantaneous eigenstate used is $\ket{E_-}$ in 
Eqs.~(\ref{Eq:XX_eigenstates}) or~(\ref{Eq:Heisen_eigenstates}), leading to slight differences 
between panels (a) and (b). In panels (b) and (c), we see that the ``up" state ($\ket{0}$) 
is transferred from the left-most qubit to the right-most qubit, following a similarly 
counter-intuitive pulse sequence.  Since the $H_3$ operators are identical for 
the subspaces ${\cal H}_{1/2}$ and ${\cal H}_{-1/2}$, their separate evolutions are also 
identical. Thus, as illustrated in Fig.~\ref{Fig1}, an arbitrary state $a\ket{0} + b\ket{1}$ 
of qubit $1$ in Eq.~(\ref{Eq:initial_state}) is transmitted to qubit $3$ via 
two mutually-coherent evolutions.

\begin{figure}[htbp]
\includegraphics[scale=0.9, angle=0]{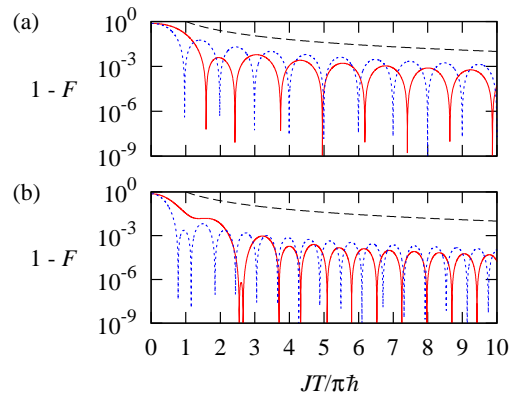}
\caption{(color online). Infidelity $(1-F)$ is plotted as a function of $JT/\pi\hbar$ for 
(a) {\it XX} and (b) Heisenberg couplings, using linear and harmonic interpolation 
functions (solid red and dotted blue lines, respectively). The black dashed line represents 
$(1-F)\propto 1/(JT)^2$ for comparison.}
\label{Fig:XY_HS_Fidelity}
\end{figure}

The adiabatic solution described above is only valid when $JT/\hbar \gg 1$. In this limit, 
Eqs.~(\ref{Eq:XX_eigenstates}) and (\ref{Eq:Heisen_eigenstates}) give a perfect (adiabatic) 
fidelity, $F_{\rm ad}(T) = 1$. When $T$ is finite however, the adiabatic theorem predicts that 
$1-F \propto (JT/\hbar)^{-2}$. To obtain an infidelity $1-F < 10^{-6}$, the adiabatic gate time 
should be $JT/\hbar \sim {\cal O}(10^3)$, \emph{much} longer than a conventional gate, for 
which $JT/\hbar\sim {\cal O}(1)$. Such slow adiabatic evolution could obviously cause problems, 
despite its intrinsic fault tolerance. However, when we perform a numerical integration of 
the time-dependent Schr\"odinger equation governed by Hamiltonian~(\ref{Eq:Hamiltonian_3}), 
we find that the infidelity $1-F$ as a function of evolution time $T$ is far from a smooth 
quadratic function. Instead, while it approaches the predicted upper envelope, there are 
also striking resonance features where the infidelity dips to zero, as shown in 
Fig.~\ref{Fig:XY_HS_Fidelity}.

\begin{figure}[htbp]
\includegraphics[scale=0.9,angle=0]{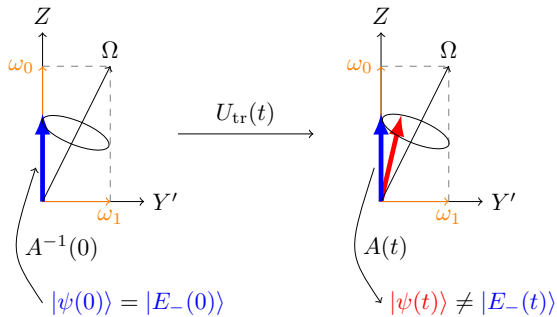}
\caption{(color online). Schematic representation of the time evolution in the 
adiabatic frame, given by Eq.~(\ref{Eq:exact}). The instantaneous eigenvector is 
represented by a thick blue arrow and the exact evolved state by a thick red arrow.}
\label{Fig:adiabatic_frame}
\end{figure}

The origin of the unexpected resonances in AQT fidelity becomes clear when we consider 
the {\it XX} coupling with harmonic interpolation functions.  For this special case, 
we can obtain an analytical solution by transforming into the adiabatic frame~\cite{Carroll}, 
as illustrated in Fig.~\ref{Fig:adiabatic_frame}. We define
\begin{align}
D(t) = A^{-1}(t)H_3(t) A(t) \,,\quad \ket{\psi(t)} = A(t)\ket{\phi(t)}\,,
\end{align}
where the column vectors of $A(t)$ are the instantaneous eigenstates given by 
Eq.~(\ref{Eq:XX_eigenstates}). The Schr\"odinger equation in the adiabatic frame 
takes the form
\begin{subequations}
\begin{align}
i\hbar\frac{\partial}{\partial t} \ket{\phi(t)}
&=\left[ D(t) -i\hbar A^{-1}(t)\frac{\partial A(t)}{\partial t}\right]
\ket{\phi(t)} \,,\\
&= H_{\rm tr}\ket{\phi(t)}\,.
\end{align}
\end{subequations}
For harmonic interpolation and {\it XX} couplings, the transformed Hamiltonian 
$H_{\rm tr}$ becomes time-independent:
\begin{subequations}
\begin{align}
H_{\rm tr} &=
2J
\begin{bmatrix}
-1 & 0 & 0\\
0  & 0 & 0\\
0  & 0 & 1\
\end{bmatrix} +
\frac{\pi\hbar}{2T}\frac{1}{\sqrt{2}}
\begin{bmatrix}
0  & i & 0 \\
-i & 0 & -i \\
0  & i & 0
\end{bmatrix}\\
&= \hbar\omega_0\,Z + \hbar\omega_1\,Y'\,,
\end{align}
\end{subequations}
where $\hbar\omega_0 \equiv 2J$ is the absolute value of the ground state energy and 
$\omega_1 \equiv {\pi}/{2T}$ is the frequency of the switching functions, 
$f(t)=\cos(\omega_1 t)$ and $g(t)=\sin(\omega_1 t)$. The matrix $Y'$, which resembles 
the angular momentum operator $I_y$ of a spin-1 system, is responsible for 
the non-adiabatic behavior. It has the same eigenvalues as $Z$, i.e., $0$ and $\pm 1$. 
The Hamiltonian in the adiabatic frame, 
$H_{\rm tr} = \hbar\Omega (Z \cos\alpha + Y' \sin\alpha)$, has the eigenvalues
\begin{subequations}
\label{Eq:Nonadiabatic_Eigen}
\begin{align}
e_0=0\,,\; e_\pm = \pm\hbar\Omega\,, \quad
\text{with}\quad \Omega \equiv \sqrt{\omega_0^2 + \omega_1^2}\,,
\end{align}
and the corresponding eigenstates
\begin{align}
\ket{e_0} =\frac{1}{\sqrt{2}}
\begin{bmatrix}
-\sin\alpha \\
i\sqrt{2}\cos\alpha \\
\sin\alpha
\end{bmatrix},\,
\ket{e_{\pm}} =
\frac{1}{2}
\begin{bmatrix}
1 \mp\cos\alpha\\
\mp i\sqrt{2}\sin\alpha\\
1\pm\cos\alpha
\end{bmatrix},
\end{align}
\end{subequations}
where $\tan\alpha \equiv {\omega_1}/{\omega_0}$.

As illustrated in Fig.~\ref{Fig:adiabatic_frame}, the time evolution in the adiabatic 
frame is analogous to the rotation of a spin-1 system around an effective, constant 
magnetic field given by ${\bf \Omega} = \omega_0 \bm{\hat{Z}} + \omega_1 \bm{\hat{Y}'}$,
where $\bm{\hat{Y}'}$ is the rotation axis associated with matrix $Y'$. The state vector 
is initially oriented along $\bm{\hat{Z}}$, which corresponds to $\ket{E_-(t)}$ in the
original frame of Eqs.~(\ref{Eq:XX_eigenstates}). In the adiabatic limit 
$\omega_0 \gg \omega_1$, the precession axis is ${\bf \hat{\Omega}} = \bm{\hat{Z}}$, 
so the state vector does not precess. Thus, in the original frame, the state vector is 
given by $\ket{E_-(t)}$ for all $t$.

Non-adiabatic evolution occurs when $\omega_1>0$.  The state vector is initially 
aligned with $\bm{\hat{Z}}$ in the adiabatic frame; however it precesses when 
${\bf \hat{\Omega}} \neq \bm{\hat{Z}}$. As the state vector deviates from 
$\bm{\hat{Z}}$ in the adiabatic frame, it also deviates from the adiabatic ground 
state $\ket{E_{-}(t)}$ in the original frame. After a full precession period given 
by $\Omega T = 2\pi n$, the state vector returns to the $\bm{\hat{Z}}$ direction, 
or the ideal target state $\ket{E_{-}(T)}$. The physical picture is analogous to 
the magnetic resonance of a spin-$1/2$ particle in a static magnetic field, with 
a small perpendicular ac field.  In this case, the state vector precesses
about a static magnetic field in the rotating frame~\cite{Cohen,Slichter}.

As depicted in Fig.~\ref{Fig:adiabatic_frame}, the evolved state in the original 
frame is given by
\begin{align}
\ket{\psi(t)} = A(t)\, U_{\rm tr}(t)\, A^{-1}(0)\, \ket{\psi(0)}\,,
\label{Eq:exact}
\end{align}
where the time evolution operator in the adiabatic frame is given by 
$U_{\rm tr} = e^{-iH_{\rm tr}t/\hbar}$. The fidelity at time $t=T$ can be obtained exactly:
\begin{align}
F(T) &=\frac{1}{4} \Bigl[\,\cos^2(\Omega T) (1+ \cos^2\alpha)^2 +
4\sin^2(\Omega T)\cos^2\alpha \nonumber\\
&+2\cos(\Omega T)\sin^2\alpha(1+\cos^2\alpha) + \sin^4\alpha \,\Bigr]\,.
\label{Eq:XX_analytic_fidelity}
\end{align}
The results are indistinguishable from the numerical solution shown in 
Fig.~\ref{Fig:XY_HS_Fidelity}(a).  Perfect fidelity occurs at the resonance condition 
$\Omega T = 2\pi n$, which is given by
\begin{align}
JT/\hbar\pi = \sqrt{n^2 -\frac{1}{16}} \approx n\,,\quad n\in \mathbb{N}\,.
\end{align}

We have now identified two paths to perfect teleportation. The first corresponds to 
the asymptotic (adiabatic) limit on the far right-hand side of Fig.~\ref{Fig:XY_HS_Fidelity}. 
The second occurs at any one of the resonant conditions. It is interesting that 
resonances only occur in certain interpolation schemes. For example, the quadratic 
interpolation $f(s) = 1 -s^2$ and $g(s) = s(2-s)$ has resonances, while $f(s) = 1
-s^2$ and $g(s) = s^2$ does not.

Our results can be tested experimentally using current technology. Controllable 
three-qubit systems have been demonstrated in quantum dots~\cite{Gaudreau12} and 
superconductors~\cite{Martinis11}. Single-shot measurements and the preparation of 
singlet states are almost routine~\cite{Buluta11}. AQT could therefore be implemented 
as follows. Qubit 1 is initially prepared in the ``up" state, while qubits 2 and {3}
are prepared in a singlet state. After switching $f$ and $g$ according to the AQT 
protocol, qubit 3 is measured. Repeating this experiment many times provides a fidelity 
estimate for AQT, over the evolution period $T$. The resonant peaks of the fidelity 
can be examined in the time domain by varying $T$. Since AQT corresponds to two 
parallel APs for the two spin components of a qubit, we could also explore interesting
phenomena like coherent population trapping and electromagnetically induced 
transparency~\cite{Scully}, which have also been studied in the context of AP, 
for three-level atoms.

So far we have only explored AQT with three qubits, where qubits $2$ and $3$ are 
initially in a singlet state -- the same initial state used for conventional quantum 
teleportation.  An interesting next step would be to study AQT over longer distances.
Our preliminary numerical studies suggest that AQT could be implemented in a more 
general spin chain geometry. We leave this for future work.

In conclusion, we have shown that adiabatic quantum teleportation consists of two 
adiabatic passages corresponding to the quantum information transfer of ``up" and 
``down" components of a qubit. When this protocol is performed non-adiabatically, 
resonances occur in the fidelity, in analogy with magnetic spin resonance. 
The observation of resonances points toward a new paradigm for fast and robust 
adiabatic gates. Our results can be tested experimentally using superconducting or 
spin qubits, with currently available technologies.  

\begin{acknowledgments}
We would like to thank J. H. Eberly for pointing out Ref.~\cite{Carroll}. This work 
was supported by the DARPA QuEST through AFOSR and NSA/LPS through ARO.
\end{acknowledgments}

\end{document}